# Identification of a Proliferation Gene Cluster Associated with HPV E6/E7 Expression Level and Viral DNA Load in Invasive Cervical Carcinoma

Christophe Rosty[1,2], Michal Sheffer[3*], Dafna Tsafrir[3*], Nicolas Stransky[2], Ilan Tsafrir[3], Martine Peter[1], Patricia de Crémoux[1], Anne de La Rochefordière[4], Rémy Salmon[5], Thierry Dorval[6], Jean Paul Thiery[7], Jérôme Couturier[1], François Radvanyi[2], Eytan Domany[3], Xavier Sastre-Garau[1]

[*]These authors have contributed equally to the analysis of the data

Département de Biologie des Tumeurs[1], Oncologie Moléculaire UMR144 CNRS[2], Département de Radiothérapie[4], Département de Chirurgie[5], Département d'Oncologie Médicale[6], Département de Transfert[7], Institut Curie, 26 rue d'Ulm, 75248 Paris Cedex 05, France

Department of Physics of Complex Systems[3], Weizmann Institute of Science, Rehovot, Israel

Proofs addressed to:   Dr. Xavier Sastre-Garau

Departement de Pathologie

Institut Curie, Section Médicale

26 rue d'Ulm

75248 Paris Cedex 05, France

Email : xavier.sastre@curie.net

Phone : +33 1 44 32 42 50

Fax : +33 1 44 32 40 72

**ABSTRACT**

Specific HPV DNA sequences are associated with more than 90% of invasive carcinomas of the uterine cervix. Viral E6 and E7 oncogenes are key mediators in cell transformation by disrupting TP53 and RB pathways. To investigate molecular mechanisms involved in the progression of invasive cervical carcinoma, we performed a gene expression study on cases selected according to viral and clinical parameters. Using Coupled Two-Way Clustering and Sorting Points Into Neighbourhoods methods, we identified a 'Cervical Cancer Proliferation Cluster' composed of 163 highly correlated transcripts, many of which corresponded to E2F pathway genes controlling cell proliferation, whereas no TP53 primary target was present in this cluster. The average expression level of the genes of this cluster was higher in tumours with an early relapse than in tumours with a favourable course (P=0.026). Moreover, we found that E6/E7 mRNA expression level was positively correlated with the expression level of the cluster genes and with viral DNA load. These findings suggest that HPV E6/E7 expression level plays a key role in the progression of invasive carcinoma of the uterine cervix via the deregulation of cellular genes controlling tumour cell proliferation. HPV expression level may thus provide a biological marker useful for prognosis assessment and specific therapy of the disease.

**INTRODUCTION**

DNA sequences of specific HPV types are detected in the vast majority of invasive cervical carcinoma (Bosch et al., 1995), a worldwide and frequent disease (Ferlay et al., 2001). HPV 16 and 18, corresponding to highly oncogenic genotypes, are detected in 59% and 15% of the cases, respectively (Clifford et al., 2003). E6 and E7 viral oncoproteins are major contributors to neoplastic progression by interfering with cell cycle G1-S checkpoint (for review, (zur Hausen, 2002)). Among a variety of cellular targets, E6 binds and degrades TP53 protein by forming a complex with the ligase E6AP, leading to genetic instability. E7 abrogates pRB protein function through its ubiquitination-mediated degradation, which leads to activation of E2F regulated genes and deregulates the progression through the G1 phase of the cell cycle. Integration of viral sequences into the host genome interrupts E2 open reading frame, leading to the constitutive expression of E6/E7 in the transformed cells (Romanczuk & Howley, 1992).

Most cases of early stage invasive cervical carcinoma can be cured by a combination of surgery, radiotherapy and chemotherapy (Gerbaulet et al., 1992; Morris et al., 1999). However, some tumours relapse at short term and are lethal in most of the cases, despite chemotherapy (Omura et al., 1997). Little is known about the biological mechanisms which could account for these differences in clinical behavior. Viro-clinical studies have reported that the outcome of cervical cancer was related to the type of HPV associated to the tumour. A favourable course was observed for tumours associated with HPV58 and related types (Lai et al., 1999) whereas association with HPV18 was found to be indicative of poor outcome (Burger et al., 1996; Lombard et al., 1998).

To get insight into the molecular mechanisms controlling the progression of invasive cervical carcinoma, we have designed a gene expression study on cases selected according to viral and clinical parameters. HPV16- and HPV18-associated tumours were included in order to determine whether specific gene expression profile could characterize these HPV types. To determine whether a characteristic pattern of gene expression could be linked to the disease course, we also analysed cases with favourable outcome and tumours which presented an early relapse uncontrolled by the treatment. A combination of unsupervised Coupled Two-Way Clustering (CTWC) (Getz et al., 2000) and Sorting Points Into Neighbourhoods (SPIN) (Tsafrir et al., 2005) methods was employed to mine the expression data, together with the use of rigourous statistical tests, thus combining the benefits of both knowledge and data driven approaches. One major finding of our analysis was the identification of a 'Cervical Cancer Proliferation Cluster' (CCPC) composed of 163 highly correlated transcripts, many of which corresponded to genes controlling cell proliferation. We found that tumours with an early relapse had an average expression level of CCPC genes higher than that of tumours with a favourable course, suggesting that the CCPC may be indicative of disease outcome. Moreover, we showed that E6/E7 mRNA expression was positively correlated with the expression level of the CCPC genes and to viral DNA load. Altogether, these findings suggest that tumour cell proliferation is dependent on E6/E7 mRNA levels and that HPV DNA load, positively correlated to E6/E7 mRNA level, may be associated with the outcome of invasive carcinoma of the uterine cervix.

## RESULTS

### Global data overview

<u>Unsupervised analysis separated tumour samples according to their histological type</u>

Gene expression profiling was performed on 45 samples (5 normal mucosa, 5 cell lines, and 35 primary tumours including 5 duplicates) with Affymetrix HG-U133A oligonucleotide microarray (Table 1). Samples in duplicates exhibited high similarity in expression profiles (average correlation of 0.95). In order to generate an overview of the data and to identify major partitions and relationships, we filtered the genes with highest variance and ordered the resulting expression matrix in SPIN (Figure 1). Two separate ordering operations were performed: one on the genes (rows; Figure 1c) and another on the samples (columns; Figure 1b). The two-way organized expression matrix (Figure 1d) permitted thus to study concurrently the structure of both samples and genes. Unsupervised ordering in the context of the most varying transcripts separated the samples in complete agreement with the nature of the 3 types of samples: normal mucosa, primary tumours and cell lines. Furthermore, a clear distinction was seen within the tumour samples according to their histological type (SCC versus AC). All this information is visually displayed in the SPIN permutated distance matrix for the samples (Figure 1b). While the PCA image (Figure 1a) provides only the top principal directions (here, 3), the distance matrices and the reordered expression matrix contain the full high-dimensional relationships (Tsafrir et al., 2005).

At this stage we did not detect any expression signal associated with differences in viral type or disease outcome.

<u>Supervised analysis</u>

Supervised hypothesis testing corroborated our observations regarding the grouping of samples. Using the t-test with 5% FDR statistical confidence, we found that 2507 of 22,215 probe sets (11.3%) were differentially expressed in the tumour samples as compared with the normal samples, with 1206 probe sets (which include 849 unique annotated genes and 178 ESTs) showing overexpression in tumour samples (supplementary tables). Among these, the major and most significant functional groups were: DNA metabolism (n=96), mitotic cell cycle (n=93), regulation of cell cycle (n=59), DNA replication and chromosome cycle (n=56), and DNA repair (n=35). Among the known genes which were overexpressed in tumours as compared to normal cervix, 94 were also found to be overexpressed in primary tumours when compared to the cell lines (supplementary tables). The majority of these genes were involved in immune response and were related to stroma cells. Among the tumour samples, 6.86% of the probe sets were differentially expressed in SCC as compared to AC.

As in unsupervised analysis, no single gene separated the tumour samples according to either viral type or disease outcome (using the constraint of 5% FDR significance level).

**A gene cluster associated with disease outcome includes mostly proliferation genes**

In a second step, we focused on one gene cluster, including 163 probe sets, identified by using CTWC on the 5000 probe sets with the highest variance. The expression profile of the genes of this cluster separated the samples into four groups: a group composed of all normal samples, a second group including 7 primary tumours among which 6 presented a favourable outcome ('favourable outcome group'), a third group containing the remaining primary tumours, and a fourth group composed of all cell lines (Figure 2). The P value for having only 6 favourable outcome tumours and 1 unfavourable outcome tumour in one group is 0.06, according to one-tail Fisher exact test. It should be stressed that this is not a separation according to outcome, since only a subset of tumours with favourable outcome belongs to the 'favourable outcome group'. This makes it impossible to identify this group of genes by using a supervised test designed according to tumour outcome. Figure 2 shows the expression matrix of the corresponding dendrogram: the normal samples had the lowest expression levels, the 'favourable outcome group' was closest to the normal samples and the cell lines had the highest expression levels.

The 163 probe sets included in this gene cluster correspond to 123 unique genes and 16 ESTs (Table 2). Looking at Gene Ontology biological process annotations, we found that 55 of these genes were related to cell cycle, 30 to nuclear division, 29 to M-phase of mitotic cell cycle, 28 to regulation of cell cycle, and 22 to DNA replication and chromosome cycle. All these annotations have p-value < $10^{-10}$ according to Fisher Exact test. Consequently, we refer to this gene cluster as the 'Cervical Cancer Proliferation Cluster' (CCPC).

Even though not a single gene separated the patients according to outcome (at FDR of 5%), the average expression levels of the CCPC genes was higher in the unfavourable outcome tumours when compared to the favourable outcome tumours (P = 0.026, t-test). This difference becomes much more significant when the average expression levels of the CCPC genes was compared between the 6 'favourable outcome tumours' and all the other tumours (P value in the $10^{-6}$ range).

**Validation of the 'Cervical Cancer Proliferation Cluster' by qRT-PCR**

Quantitative RT-PCR was performed in order to validate the microarray expression measurements and to increase the number of samples on which the prognostic value of the proliferation cluster is tested. Twenty genes were selected from the CCPC and analysed using qRT-PCR in 70 samples: the 5 cells lines and the 28 of 30 invasive carcinomas with known disease outcome previously analysed by Affymetrix array, 30 additional invasive carcinomas and 2 additional cell lines (IC4 and IC8) (see supplementary table 1 for tumor characteristics). To select these genes, probe sets from the CCPC were sorted according to their ability to separate the 6 tumours with favourable outcome from the 13 tumours with unfavourable outcome using a T-test, and fold-change ratio between the average expressions of the two groups. The 20 genes that showed the best combination of low p-value and high fold-change were selected (Table 3). A high Pearson correlation between qRT-PCR gene expression level and Affymetrix signal for the corresponding probe set was observed (mean = 0.88, median = 0.89) (Table 3 and supplementary figure 1).

Cluster analysis of the samples, using the qRT-PCR results revealed a tumour group that contained 9 favourable outcome tumours including the 6 tumours previously identified ('favourable outcome group'), as well as 1 normal sample, and 3 unfavourable outcome tumours. The P value for having 9 favourable outcome tumours and 3 unfavourable outcome tumours in one group is 0.076 according to one-tail Fisher exact test. These results indicated that some of the genes chosen for the qRT-PCR could be potential markers for cervical cancer outcome.

**E6 and E7 expression correlates with the 'Cervical Cancer Proliferation Cluster' expression level and with viral DNA load**

The viral proteins E6 and E7 bind to and inhibit TP53 and pRB, respectively, driving the cell into proliferation (Figure 3). We therefore hypothesized that variation in expression level of the CCPC genes might correlate with E6/E7 mRNA levels. E6 and E7 mRNA expression was measured by qRT-PCR for HPV16 tumours (n = 35) and HPV18 tumours (n = 18) separately. Quantitative RT-PCR showed great variations in E7 expression levels among tumour samples. $2^{-\Delta\Delta CT}$ expression values ranged from 0.002 to 12.46 (mean 2.75±2.58) in HPV16 tumours and from 0.22 to 9.77 (mean 1.54±2.14) in HPV18 tumours. E6 expression was highly correlated with E7 expression in HPV16/18 tumours (R = 0.792, p<0.0001, linear regression). We thus used only E7 expression for further correlation analysis.

Considering the large variations in E7 expression among the different samples, we hypothesized that the mRNA levels of E7 depended on the number of HPV genomes per neoplastic cells. To test this hypothesis, E7 DNA load was also measured by qRT-PCR, for HPV16 tumours (n = 34) and HPV18 tumours (n = 17) seperately. We found that E7 mRNA expression level was correlated with E7 DNA load (Spearman correlation of 0.47 for HPV16 tumours, 0.66 for HPV18 tumours). Furthermore, E7 RNA level was highly correlated with the expression level of the CCPC genes measured by Affymetrix array and with that of the 20 selected genes measured by qRT-PCR (Table 4 and Figure 3). We also found a correlation between the expression level of the CCPC genes and E7 DNA levels although the correlation coefficient was lower than for E7 RNA levels (r = 0.33 for HPV16 tumors and r = 0.55 for HPV18 tumors) (Table 4 and Figure 3).

To evaluate whether high correlation with E7 mRNA expression was characteristic of the CCPC genes, Spearman's Rho correlation was measured between the mRNA expression levels of E7 and the expression levels of all probe sets of the HG-U133A Affymetrix microarray (supplementary tables and supplementary figure 2). In HPV16 tumours, 195 probe sets were highly positively correlated (R>0.7) to E7 mRNA expression levels while 230 probe sets were found correlated in HPV18 tumours. CCPC genes were over-represented among these probe sets, both in HPV16 tumours (55/195, 33.7%) and in HPV18 tumours (37/230, 22.7%). A group of 33 probe sets had a correlation coefficient >0.7 for both HPV16 and HPV18 tumours ($P < 10^{-30}$, hypergeometric test). Gene ontology annotations of these probe sets are cell cycle, M-phase, S-phase, regulation of cell cycle, DNA regulation and DNA metabolism. From the 20 known genes of these 33 probe sets, 11 (55%) are known targets of the RB-E2F pathway (Markey et al., 2002; Muller et al., 2001; Ren et al., 2002). Surprisingly, genes negatively correlated to E6/E7 expression levels did not include many known TP53 targets, such as CDKN1A or GADD45A.

**Correlation between gene expression analysis and disease outcome**

Assigning an outcome indicator value -1 for unfavourable prognosis and +1 for favourable prognosis, we measured the Pearson correlation of various quantities with outcome. Pearson correlation coefficient between the average expression level of the 163 CCPC probes sets and outcome was 0.42 (P = 0.02). Interestingly, this correlation coefficient was higher for the HPV18 tumours compared to the HPV16 tumours: R = 0.65 (P = 0.01) versus R = 0.31 (P = 0.1), respectively. We also calculated Pearson correlation coefficient between disease outcome and E7 mRNA expression levels. We obtained a significant correlation only for the HPV18 tumours: R = 0.39 (P = 0.01).

**DISCUSSION**

Gene expression profiling in cervical carcinoma specimens using CTWC analysis identified a cluster of 163 transcripts, mostly related to cell proliferation (CCPC genes) and differentially expressed according to disease outcome. Importantly, expression levels of the CCPC genes were found positively correlated to E6/E7 mRNA levels. These results indicate that, in agreement with the observations performed by *in vitro* studies on cell lines, E6 and E7 viral oncogenes play a key role in the progression of invasive cervical carcinoma via the deregulation of cellular genes controlling cell proliferation. Interestingly, 54 of the 123 CCPC genes (44%) correspond to previously reported E2F targets (Table 2) (Markey et al., 2002; Muller et al., 2001; Ren et al., 2002). The mechanisms by which E7 interferes with the regulation of proliferation, notably the inactivation of the pRB protein and the subsequent release of active E2F transcription factor, have been largely documented (Munger et al., 2004). It is of interest to stress that analyses of gene expression profiles in cancer-derived cells have identified clusters of genes under the control of E2F, which are largely common to the CCPC genes (Milyavsky et al., In press; Thierry et al., 2004; Wells et al., 2003). E2F is in turn controlled directly by RB and indirectly by TP53 (Tabach et al., Submitted). In HPV18-associated HeLa cells, the switch of E6/E7, through the regulated expression of E2, led Thierry *et al.* to identify a subset of 28 mitotic genes, among which 19 (68%) were common to the CCPC genes and 12 corresponded to E2F targets (AURKB, CDC20, CCNA2, CCNB2, MAD2L1, MKI67, MYBL2, NEK2, PTTG1, RRM2, TOP2A, UBE2C) (Thierry et al., 2004). Large-scale gene expression analyses of cervical neoplasia published so far aimed at identifying molecular markers associated with the progression of lesions (Chen et al., 2003; Sopov et al., 2004; Wong et al., 2003) but reported no data concerning the expression of the viral oncogenes. Chen *et al.* identified 62 genes overexpressed in high-grade compared to low-grade squamous intra-epithelial lesions, four of which were common to the CCPC genes (TK1, MYBL2, MCMC4, TOP2A). In our study, the E6 and E7 genes were found to be co-expressed in tumour samples and the respective impact of these viral oncoproteins on cell proliferation has to be specified. However, almost no primary known TP53 targets were found negatively correlated to E6/E7 expression levels among all Affymetrix probe sets. This result suggest that the expression level of E7 is the main driver of tumor cell proliferation in cervical cancer. In contrast, possibly, a high sensitivity of TP53 to even low levels of E6 may be sufficient to inactivate the TP53 pathway.

Another striking feature of our results was that a wide range of E6/E7 expression levels was observed among tumour cases. A putative role for cellular genes on the control of viral oncogenes has not been reported in cancer cells and we hypothesized that differences in E6/E7 mRNA expression levels could be related, at least in part, to variations in HPV DNA copy number between tumours. We found a positive correlation between E6/E7 mRNA levels and viral DNA load in tissue specimens. Histological analysis showed that all samples contained more than 50% of invasive carcinoma cells and it is unlikely that the correlation observed was related to differences in carcinoma cell density or in tumour differentiation.

The differences in E6/E7 expression levels observed between cases may also be related to differences in the physical state of viral DNA. Although integrated sequences are detected in most invasive carcinoma (Cullen et al., 1991), a proportion of cases contains only episomal HPV DNA (Klaes et al., 1999; Matsukura et al., 1989). A deregulated expression of integrated viral DNA (Schneider-Gadicke & Schwarz, 1986) and a high stability of E6/E7 mRNAs derived from integrated sequences (Jeon et al., 1995) account for a higher level of E6/E7 expression derived from integrated viral DNA compared to that from episomal viral DNA. It is worth to be mentioned that most of HPV18 sequences associated with cervical cancer are found integrated into the host genome (Cullen et al., 1991). A high transcription rate of HPV18 oncogenes can thus account for the rapid progression of cervical carcinoma associated with this virus type (Lombard et al., 1998; Wang & Lu, 2004).

Viro-clinical analyses reported that a high viral load was related to a higher risk of progression of low-grade to high-grade intraepithelial neoplasia (Dalstein et al., 2003) and to the persistence of high-grade lesions (Ho et al., 1995), but few studies have analysed the influence of HPV DNA copy number on the course of invasive cancers. A high HPV DNA load has been found positively correlated to tumour differentiation (Ikenberg et al., 1994) and negatively correlated to clinical stage (Berumen et al., 1994). However, no link has been reported between viral DNA load and cell proliferation or disease outcome. Analysis of a large number of cases at different clinical stages is needed to determine whether viral DNA load could be used as an independent biological marker of the outcome of invasive carcinoma. To further support the correlation with disease outcome, we checked the CCPC on an available breast cancer gene expression dataset (van 't Veer et al., 2002). This study was based on a different DNA microarray, in which only 49 genes corresponding to the 163 probe sets of the CCPC were represented. Among those, we searched for genes that could separate unfavourable outcome tumours from favourable outcome tumours. Using t-test, we found that the expression levels of 31 genes passed the 5% FDR threshold and were able to separate favourable outcome breast tumours from unfavourable outcome tumours (supplementary figure 3). These results indicate that the CCPC may be useful to predict outcome in other tumour types.

In summary, CTWC analysis in a series of invasive carcinoma of the uterine cervix identified a proliferation gene cluster whose expression was found positively correlated with E6/E7 viral oncogenes expression and, to a lower extent, with HPV DNA load. HPV expression level may thus correspond to a biological marker of interest in both prognosis assessment and targeted therapy design of invasive carcinoma of the uterine cervix.

**MATERIALS AND METHODS**

**Cervical Tissue Samples and Cell Lines**

Primary tumour samples (n=60) from invasive cervical carcinoma were selected from the Institut Curie tumour bank. Characteristics are listed in table 1 and supplementary table 1. Patient's age ranged from 23 to 78 year-old (median, 46). HPV status was determined by PCR, as described (Rosty et al., 2004). Relapse-free survival (RFS) was defined as the interval elapsed between the date of the first symptoms and that of local recurrence and/or distant metastasis. Cases with RFS >5 years (n=30) were classified as diseases with favourable outcome and those with RFS <3 years (n=27) as diseases with unfavourable outcome.

All tumour samples had been flash-frozen and stored at -80°C. Histological analysis of tumour tissues adjacent to the selected samples showed that tumour samples contained >50% of invasive carcinoma cells.

Normal exocervical mucosa has been sampled and flash-frozen from 5 hysterectomy specimens, removed for non-cervical diseases. Cell lines (IC1, IC3 to 8) were established from human primary invasive cervical carcinoma (Couturier et al., 1991; Sastre-Garau et al., 2000). Samples from the primary tumours corresponding to IC5, IC6, and IC8 were included in this series (#25, #16, and #31, respectively).

**Labeling and Microarray Hybridization**

A total of 45 cervical samples were analysed: 30 taken from invasive carcinoma with 5 duplicates, 5 from normal mucosa, and 5 corresponding to carcinoma-derived cell lines (IC1, IC3, IC5 to 7) (Table 1). Total RNAs were extracted from each sample by caesium chloride ultracentrifugation. RNA quality was assessed by visualization of the 28S/18S ribosomal RNA ratio on electrophoresis gel. Complementary RNA target was prepared and labelled as described in the Affymetrix GeneChip Expression Analysis Technical Manual (High

Wycombe, United Kingdom). The labelled target was hybridized to Affymetrix HG-U133A oligonucleotide microarray, representing 22 215 probe sets. To control the reproducibility of the results, hybridization was performed in duplicate for 5 tumour samples.

**Quantitative Real-Time PCR**

Reverse transcription was performed using 1 µg of total RNA, random hexamer primer, and the SuperScript II reverse transcription kit (Invitrogen, Cergy Pontoise, France). Real-time PCR was performed in the SYBR Green format (Applied Biosystems, Courtaboeuf, France) for STK6, H2AFZ, KPNA2, CDC20 and to amplify E6 and E7 HPV transcripts for HPV16 tumours and HPV18 tumours. For normalization, TBP expression was used. Primer sequences were GTC AGT ACA TGC TCC ATC TTC (forward) and GTG AAT TCA ACC CGT GAT ATT C (reverse) for STK6; CTC ACC GCA GAG GTA CTT G (forward) and TTG TCC TTT CTT CCC AAT CAG (reverse) for H2AFZ ; TCA AGC TGC CAG GAA ACT ACT (forward) and GCC TTG GTT TGT TCT GAT GTC (reverse) for KPNA2; CTG TCC AGT GGT TCA CGT TC (forward) and CCT TGA CAG CCC CTT GAT G (reverse) for CDC20 ; GAG CGA CCC AGA AAG TTA CCA (forward) and AAA TCC CGA AAA GCA AAG TCA (reverse) for E6 HPV16; TCC AGC TGG ACA AGC AGA AC (forward) and CAC AAC CGA AGC GTA GAG TC (reverse) for E7 HPV16; AAT AAG GTG CCT GCG GTG (forward) and CTT GTG TTT CTC TGC GTC GT (reverse) for E7 HPV16; AAC ATT TAC CAG CCC GAC GA (forward) and TCG TCT GCT GAG CTT TCT AC (reverse) for E7 HPV18; AGT GAA GAA CAG TCC AGA CTG (forward) and CCA GGA AAT AAC TCT GGC TCA T (reverse) for TBP.

The Applied Biosystems Assays-on-Demand™ Gene Expression system (Applied Biosystems, Courtaboeuf, France) was used to analyse gene expression of 16 human genes: ANKT, GGH, CCNB2, BUB1B, FEN1, CCNB1, OIP5, MELK, MCM4, UBE2C, PLK, CDC2, ZWINT, CCNA2, TOPK, RRM2. All samples were tested in duplicate. Analysis was performed using SDS v2.1 software (Applied Biosystems) according to the manufacturer's instructions. For each mRNA sample, a difference in $C_T$ values ($\Delta C_T$) was calculated by taking the mean $C_T$ of duplicate reaction and subtracting the mean $C_T$ of the duplicate reaction of the reference (TBP) RNA. A normal cervical sample was used as the calibrator. The $2^{-\Delta\Delta C_T}$ method was used for quantification of gene expression (Livak & Schmittgen, 2001).

HPV viral load was quantified on tumour DNA using E7 primers specific for HPV16 and HPV18, in 34 HPV16 tumours and in 17 HPV18 tumours (same primer sets as for RNA expression analysis). We chose PSA as the reference gene. Primers for PSA were AGG CTG GGG CAG CAT TGA AC (forward) and CAC CTT CTG AGG GTG AAC TTG (reverse). Comparative genomic hybridization analysis of the same tumour samples (manuscript in preparation) showed that the chomosomal location where PSA maps (19q13) has little if any variation in DNA copy number.

**Data Analysis**

Data Preprocessing. The Microarray Suite 5.0 software (MAS v5.0, Affymetrix) was used to scale the raw data and produce an expression matrix, where each value was the expression level of one transcript measured in one sample. In order to avoid working with unreliably small numbers, gene expression values below 10 were set to 10 and a log2 transformation was applied (Tsafrir et al., 2005). When the Affymetrix data was used to measure correlation with E7 PCR measurements, log2 transformation and thresholding were not applied.

Genes were chosen for unsupervised analysis on the basis of their standard deviations. For duplicated tumour samples, the assigned value was the average of the two duplicates (except for the unsupervised analysis in the global overview, where both duplicates were represented).

Unsupervised analysis: Since hypothesis testing can not reveal unexpected partitions, unsupervised techniques, such as clustering, are more suited for such a task. The CTWC method (Getz et al., 2000) focuses on correlated subsets of genes and samples, such that when one is used to cluster the other, stable and significant partitions emerge. The underlying algorithm is based on iterative clustering, enabling identification of biologically relevant subsets of the data. This reveals partitions and correlations that are masked when the full dataset is used in the analysis. For example, when a particular set of genes is used to cluster the samples we find that they divide into 2 groups: a relatively tight cluster of predominantly favourable outcome tumours, and a larger cluster containing both favourable and unfavourable outcome tumours. The statistical significance of this 'favourable outcome group' was measured with Fisher exact test (Fisher, 1935).

Another exploratory analysis method that uses groups of correlated genes for meaningful ordering of tumours is SPIN (Tsafrir et al., 2005), our recently proposed methodology for data organization and visualization. At the heart of this method is a presentation of the full pairwise distance matrix of the samples, viewed in pseudo-color. The samples are iteratively permuted in search of an optimal ordering, i.e. one that can be used to study embedded shapes. Hence, certain structures in the data (elongated, circular and compact) manifest themselves visually in a SPIN generated distance matrix.

<u>Supervised analysis</u>. Supervised methods were employed in order to expand and refine the list of genes that was obtained by the unsupervised step. In order to control contamination with false positive genes associated with multiple comparisons, we used the method of Benjamini and Hochberg (Benjamini & Hochberg, 1995) that defines the average false discovery rate (FDR); namely, the fraction of false positives among the list of differentiating genes.

<u>qRT-PCR analysis</u>

The data analysis of the qRT-PCR for the selected genes was based on samples previously analysed with Affymetrix array with 30 additional primary tumours and 2 additional cell lines. Missing values were completed using a K-nearest neighbours algorithm (Troyanskaya et al., 2001).

E7 analysis

The samples used for this qRT-PCR analysis were 35 HPV16 tumours (including 2 cell lines), 16 of which were used previously for the Affymetrix arrays, and 18 HPV18 tumours (including 4 cell lines), of which 14 were used for the Affymetrix arrays. Normal samples were added with assigned values of 0 for E7 mRNA and DNA expression. Correlations of E7 expression with other genes were calculated using Spearman's Rho correlation.

Gene Ontology Annotation

For Gene Ontology (GO) annotation we used the web site http://apps1.niaid.nih.gov/David/upload.asp (Dennis et al., 2003) that produces p-values according to Fisher exact test for the statistical significance of the measured over (or under) representation of a specific functional annotation, among members of a particular group of probe sets. Another web site used for GO annotation is the Affymetrix Analysis Center http://www.affymetrix.com/analysis/netaffx/index.affx (Liu et al., 2003).

**ACKNOWLEDGMENTS**

The work at the Institut Curie was supported by the "Programme Génomique du Ministère chargé de la Recherche" for the Affymetrix experiments, La Ligue Contre le Cancer and the Association pour la Recherche contre le Cancer.

The work at the Weizmann Institute was supported in part by a grant from the Ridgefield Foundation. The work of ED at the Institut Curie was supported by a Rotschild-Mayent Fellowship.

We thank Rachida Bouarich and Jérôme Bourdin for excellent technical assistance.

TABLES

**Table 1**. Clinical and pathological characteristics of the 40 cervical tissue specimens analysed by Affymetrix HG-U133A oligonucleotide microarray. The characteristics of all specimens used are listed in supplementary table 1.

| Tissue specimens | HPV type | | | Outcome | | |
|---|---|---|---|---|---|---|
| | HPV16 | HPV18 | Others | Favourable | Unfavourable | NA |
| Primary Tumors | 16 | 12 | 2 | 15 | 13 | 2 |
| SCC (n=20) | 12 | 6 | 2 | 9 | 10 | 1 |
| AC (n=10) | 4 | 6 | 0 | 6 | 3 | 1 |
| Cell lines (n=5) | 2 | 3 | 0 | - | - | - |
| Normal cervix (n=5) | - | - | - | - | - | - |

SCC: Squamous Cell Carcinoma; AC: Adenocarcinome; NA: Non Available.

**Table 2**. List of the 123 unique known genes corresponding to the 163 transcripts of the "Cervical Cancer Proliferation Cluster", identified by coupled two-way clustering analysis. Genes previously reported as E2F targets are underlined.

| Probe Set ID | Gene Symbol | Gene Title |
|---|---|---|
| 212186_at | ACACA | acetyl-Coenzyme A carboxylase alpha |
| 218039_at | ANKT | <u>nucleolar protein ANKT</u> |
| 208103_s_at | ANP32E | acidic nuclear phosphoprotein 32 family, member E |
| 206632_s_at | APOBEC3B | apolipoprotein B mRNA editing enzyme, catalytic polypeptide-like 3B |
| 218115_at | ASF1B | <u>ASF1 anti-silencing function 1 homolog B</u> |
| 204244_s_at | ASK | <u>activator of S phase kinase</u> |
| 219918_s_at | ASPM | asp (abnormal spindle)-like, microcephaly associated |
| 209464_at | AURKB | <u>aurora kinase B</u> |
| 202094_at | BIRC5 | baculoviral IAP repeat-containing 5 (survivin) |
| 204531_s_at | BRCA1 | <u>breast cancer 1, early onset</u> |
| 212949_at | BRRN1 | barren homolog |
| 209642_at | BUB1 | BUB1 budding uninhibited by benzimidazoles 1 homolog |
| 203755_at | BUB1B | <u>BUB1 budding uninhibited by benzimidazoles 1 homolog beta</u> |
| 209301_at | CA2 | carbonic anhydrase II |
| 203418_at | CCNA2 | <u>cyclin A2</u> |
| 214710_s_at | CCNB1 | <u>cyclin B1</u> |
| 202705_at | CCNB2 | <u>cyclin B2</u> |
| 205034_at | CCNE2 | <u>cyclin E2</u> |
| 204826_at | CCNF | <u>cyclin F</u> |
| 203213_at | CDC2 | <u>cell division cycle 2, G1 to S and G2 to M</u> |
| 202870_s_at | CDC20 | <u>CDC20 cell division cycle 20 homolog</u> |
| 203967_at | CDC6 | CDC6 cell division cycle 6 homolog |
| 221436_s_at | CDCA3 | cell division cycle associated 3 |
| 221520_s_at | CDCA8 | cell division cycle associated 8 |
| 207039_at | CDKN2A | cyclin-dependent kinase inhibitor 2A |
| 205165_at | CELSR3 | cadherin, EGF LAG seven-pass G-type receptor 3 |
| 204962_s_at | CENPA | centromere protein A, 17kDa |
| 205046_at | CENPE | centromere protein E, 312kDa |
| 207828_s_at | CENPF | centromere protein F, 350/400ka |
| 204775_at | CHAF1B | chromatin assembly factor 1, subunit B |
| 205394_at | CHEK1 | CHK1 checkpoint homolog |
| 204170_s_at | CKS2 | CDC28 protein kinase regulatory subunit 2 |
| 202532_s_at | DHFR | dihydrofolate reductase |
| 203764_at | DLG7 | discs, large homolog 7 |
| 213647_at | DNA2L | DNA2 DNA replication helicase 2-like |
| 220668_s_at | DNMT3B | DNA (cytosine-5-)-methyltransferase 3 beta |
| 218567_x_at | DPP3 | dipeptidylpeptidase 3 |
| 217901_at | DSG2 | desmoglein 2 |
| 203270_at | DTYMK | deoxythymidylate kinase |
| 202779_s_at | E2-EPF | ubiquitin carrier protein |
| 204947_at | E2F1 | E2F transcription factor 1 |
| 202735_at | EBP | emopamil binding protein |
| 219787_s_at | ECT2 | epithelial cell transforming sequence 2 oncogene |
| 221539_at | EIF4EBP1 | eukaryotic translation initiation factor 4E binding protein 1 |
| 204817_at | ESPL1 | extra spindle poles like 1 |
| 203358_s_at | EZH2 | enhancer of zeste homolog 2 |
| 218875_s_at | FBXO5 | F-box only protein 5 |
| 204767_s_at | FEN1 | flap structure-specific endonuclease 1 |
| 202580_x_at | FOXM1 | forkhead box M1 |
| 203560_at | GGH | gamma-glutamyl hydrolase |
| 218350_s_at | GMNN | geminin, DNA replication inhibitor |
| 204318_s_at | GTSE1 | G-2 and S-phase expressed 1 |
| 205436_s_at | H2AFX | H2A histone family, member X |
| 200853_at | H2AFZ | H2A histone family, member Z |
| 218663_at | HCAP-G | chromosome condensation protein G |
| 204162_at | HEC | highly expressed in cancer, rich in leucine heptad repeats |
| 220085_at | HELLS | helicase, lymphoid-specific |
| 206074_s_at | HMGA1 | high mobility group AT-hook 1 |
| 208808_s_at | HMGB2 | high-mobility group box 2 |
| 207165_at | HMMR | hyaluronan-mediated motility receptor |
| 217755_at | HN1 | hematological and neurological expressed 1 |
| 204444_at | KIF11 | kinesin family member 11 |
| 206364_at | KIF14 | kinesin family member 14 |
| 218755_at | KIF20A | kinesin family member 20A |
| 204709_s_at | KIF23 | kinesin family member 23 |
| 209408_at | KIF2C | kinesin family member 2C |
| 218355_at | KIF4A | kinesin family member 4A |
| 209680_s_at | KIFC1 | kinesin family member C1 |
| 219306_at | KNSL7 | kinesin-like 7 |
| 201088_at | KPNA2 | karyopherin alpha 2 |
| 203276_at | LMNB1 | lamin B1 |
| 208433_s_at | LRP8 | low density lipoprotein receptor-related protein 8 |
| 202736_s_at | LSM4 | LSM4 homolog, U6 small nuclear RNA associated |
| 203362_s_at | MAD2L1 | MAD2 mitotic arrest deficient-like 1 |
| 210059_s_at | MAPK13 | mitogen-activated protein kinase 13 |
| 220651_s_at | MCM10 | MCM10 minichromosome maintenance deficient 10 |
| 202107_s_at | MCM2 | MCM2 minichromosome maintenance deficient 2 |
| 212141_at | MCM4 | MCM4 minichromosome maintenance deficient 4 |
| 204825_at | MELK | maternal embryonic leucine zipper kinase |
| 212020_s_at | MKI67 | antigen identified by monoclonal antibody Ki-67 |
| 205235_s_at | MPHOSPH1 | M-phase phosphoprotein 1 |
| 221437_s_at | MRPS15 | mitochondrial ribosomal protein S15 |
| 201710_at | MYBL2 | v-myb myeloblastosis viral oncogene homolog (avian)-like 2 |
| 204641_at | NEK2 | NIMA (never in mitosis gene a)-related kinase 2 |
| 218888_s_at | NETO2 | neuropilin (NRP) and tolloid (TLL)-like 2 |
| 213599_at | OIP5 | Opa-interacting protein 5 |
| 203228_at | PAFAH1B3 | platelet-activating factor acetylhydrolase, isoform Ib |
| 201202_at | PCNA | proliferating cell nuclear antigen |
| 204146_at | PIR51 | RAD51-interacting protein |
| 212858_at | PKMYT1 | protein kinase, membrane associated tyrosine/threonine 1 |
| 218644_at | PLEK2 | pleckstrin 2 |
| 202240_at | PLK | polo-like kinase |
| 213226_at | PMSCL1 | polymyositis/scleroderma autoantigen 1, 75kDa |
| 204441_s_at | POLA2 | polymerase (DNA-directed), alpha (70kD) |
| 213007_at | POLG | polymerase (DNA directed), gamma |
| 207746_at | POLQ | polymerase (DNA directed), theta |
| 218009_s_at | PRC1 | protein regulator of cytokinesis 1 |
| 218782_s_at | PRO2000 | PRO2000 protein |
| 203554_x_at | PTTG1 | pituitary tumor-transforming 1 |
| 222077_s_at | RACGAP1 | Rac GTPase activating protein 1 |
| 218585_s_at | RAMP | RA-regulated nuclear matrix-associated protein |
| 209507_at | RPA3 | replication protein A3, 14kDa |
| 201890_at | RRM2 | ribonucleotide reductase M2 polypeptide |
| 219493_at | SHCBP1 | likely ortholog of mouse Shc SH2-domain binding protein 1 |
| 205339_at | SIL | TAL1 (SCL) interrupting locus |
| 218653_at | SLC25A15 | solute carrier family 25 member 15 |
| 218237_s_at | SLC38A1 | solute carrier family 38, member 1 |
| 213253_at | SMC2L1 | SMC2 structural maintenance of chromosomes 2-like 1 |
| 201663_s_at | SMC4L1 | SMC4 structural maintenance of chromosomes 4-like 1 |
| 203145_at | SPAG5 | sperm associated antigen 5 |
| 204092_s_at | STK6 | serine/threonine kinase 6 |
| 218308_at | TACC3 | transforming, acidic coiled-coil containing protein 3 |
| 202338_at | TK1 | thymidine kinase 1, soluble |
| 203432_at | TMPO | thymopoietin |
| 217733_s_at | TMSB10 | thymosin, beta 10 |
| 201291_s_at | TOP2A | topoisomerase (DNA) II alpha 170kDa |
| 219148_at | TOPK | T-LAK cell-originated protein kinase |
| 210052_s_at | TPX2 | TPX2, microtubule-associated protein homolog |
| 204033_at | TRIP13 | thyroid hormone receptor interactor 13 |
| 204822_at | TTK | TTK protein kinase |
| 202589_at | TYMS | thymidylate synthetase |
| 202954_at | UBE2C | ubiquitin-conjugating enzyme E2C |
| 204026_s_at | ZWINT | ZW10 interactor |

**Table 3.** Selection of genes from the proliferation cluster genes for qRT-PCR analysis. The third column contains P values of the T-test used for comparison of expression levels between the 'favourable outcome tumour group' and the 13 unfavourable outcome tumours. The fourth column presents the fold change ratio between the average expression of the two tumour groups, and the last column shows the Pearson correlation coefficient between expression levels measured by Affymetrix array and qRT-PCR. These probe sets were selected out of 84 probe sets that passed the T-test at 5% FDR.

| Probe Set ID | Gene Symbol | P-value | Fold-change | Pearson correlation |
|---|---|---|---|---|
| 208079_s_at | STK6 | $4.99.10^{-7}$ | 2.83 | 0.93 |
| 218039_at | ANKT | $6.35.10^{-6}$ | 2.06 | 0.96 |
| 203560_at | GGH | $5.18.10^{-6}$ | 2.27 | 0.91 |
| 212141_at | MCM4 | $7.01.10^{-6}$ | 2.51 | 0.82 |
| 202705_at | CCNB2 | $7.03.10^{-6}$ | 2.54 | 0.90 |
| 204767_s_at | FEN1 | $8.81.10^{-6}$ | 1.64 | 0.89 |
| 200853_at | H2AFZ | $1.66.10^{-5}$ | 2.11 | 0.77 |
| 204825_at | MELK | $2.50.10^{-5}$ | 2.40 | 0.91 |
| 214710_s_at | CCNB1 | $2.88.10^{-5}$ | 2.91 | 0.93 |
| 209773_s_at | RRM2 | $2.95.10^{-5}$ | 2.63 | 0.94 |
| 202954_at | UBE2C | $3.39.10^{-5}$ | 2.07 | 0.80 |
| 202240_at | PLK | $6.61.10^{-5}$ | 2.14 | 0.89 |
| 202870_s_at | CDC20 | $1.61.10^{-4}$ | 2.69 | 0.94 |
| 211762_s_at | KPNA2 | $1.79.10^{-4}$ | 1.96 | 0.93 |
| 203418_at | CCNA2 | $2.19.10^{-4}$ | 2.40 | 0.88 |
| 219148_at | TOPK | $3.25.10^{-4}$ | 2.55 | 0.89 |
| 203213_at | CDC2 | $3.42.10^{-4}$ | 2.14 | 0.88 |
| 204026_s_at | ZWINT | $4.36.10^{-4}$ | 2.03 | 0.89 |
| 213599_at | OIP5 | $1.37.10^{-3}$ | 2.93 | 0.62 |
| 203755_at | BUB1B | $1.61.10^{-3}$ | 1.56 | 0.94 |

**Table 4**. Summary of the Spearman's Rho average correlation between E7 mRNA expression, E7 DNA load and the 'Cervical Cancer Proliferation Cluster', for both the Affymetrix data and qRT-PCR data. To perform the correlation calculation, normal samples were included (E7 expression was set to 0).

| Method | Tumours | Mean correlation with E7 mRNA | Mean correlation with E7 DNA |
|---|---|---|---|
| qRT-PCR | HPV16 tumours | 0.55 | 0.33 |
| | HPV18 tumours | 0.67 | 0.55 |
| Affymetrix | HPV16 tumours | 0.63 | 0.54 |
| | HPV18 tumours | 0.68 | 0.56 |

**FIGURE LEGENDS**

**Figure 1.** Global data overview. The data presented in this figure includes the 1,000 genes with highest variance over the 45 samples. Focusing on the most relevant genes by means of variance filtration facilited the computations and still gave a good overall picture of the layout of the data. (a) Projection of the samples onto the first (x-axis), second (y-axis) and third (z-axis) principal components (PC), calculated in gene-space. The nature of samples is indicated by color: normal cervical mucosa (black); carcinoma cell lines (green); primary squamous cell carcinoma (SCC, yellow) and adenocarcinoma (AC, magenta). The first PC is dominated by the differences between the cell lines and all other samples. The second PC is dominated by the differences between SCC and AC tumours. The third PC is dominated by the differences between normal samples and tumours. (b) SPIN-ordered distance matrix for the samples. Colors in the distance matrix depict dissimilarity levels between points, with red (blue) indicating large (small) distances. Hence, clusters of highly similar samples are manifested as bluish squares around the main diagonal. Note that the cell lines are a distinct, homogeneous group (marked by green in the colored bar on the right), while the normal cervical samples are clearly separated but are rather heterogeneous. (c) SPIN-ordered distance matrix for the genes. Note the grouping into several distinctive profiles. (d) Two-way SPIN-ordered expression matrix. Here colors depict relative expression intensities after centering and normalization of genes (rows), where red (blue) denotes relatively high (low) expression. Rows represent genes and columns represent samples. The colored bar below the matrix provides the tissues clinical identity.

**Figure 2.** Pattern of expression of the 'Cervical Cancer Proliferation Cluster'(CCPC). (a) The dendrogram above the matrix represents the clusters of the samples identified by CTWC. (b) In the expression matrix the samples were ordered according to the dendrogram (generated by clustering), whereas the genes were ordered by SPIN. The samples included: 5 normals, 5 cell

lines, 15 favourable outcome tumours, 13 unfavourable outcome tumours, 2 unknown outcome tumours. The color bar below the matrix displays the different origin of samples: primary tumours with unfavourable outcome (blue), favourable outcome (red), unknown outcome (cyan); cell-line (green), normal samples (black). (c) Plot of the average expression levels of the CCPC genes; the error bars indicate one standard deviation. (d) PCA diagram. Data is after log, centering and normalization. The six tumours closest to the normal samples are all with favourable outcome.

**Figure 3.** Correlations of qRT-PCR genes with E7 mRNA expression and E7 DNA load. (a) Schematic drawing of the network that controls expression of the CCPC genes, indicating the manner in which the viral proteins E6 and E7 affect the network. (b) For each gene (including E7 mRNA and DNA) the measured qRT-PCR values were ranked, separately for HPV16 and HPV18 tumours. The resulting "rank matrix" was used to sort the samples; it is presented on the left for HPV16 tumours and on the right for HPV18 tumours. Rows represent genes and columns represent samples; the entry in row *g* and column *s* represents the color code for the rank of the expression level of gene *g* in sample *s*; blue entries denote low rank and red high rank. For the sake of clarity, the ranks of the samples according to E7 mRNA and DNA measurements are also represented in the two graphs at the top and bottom. The color bar at the very bottom displays the different labels for samples, using the same color scheme as in figure 2. Spearman's Rho mean correlation with E7 mRNA and DNA levels, as described in table 4, is presented separately for HPV16 tumours and HPV18 tumours.

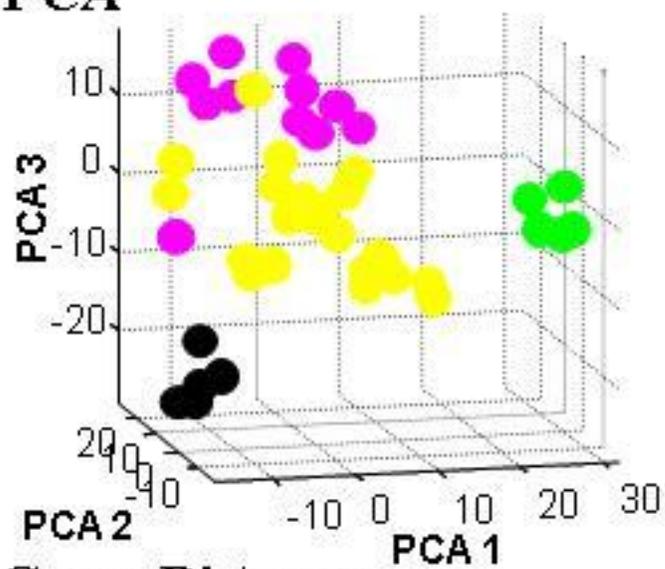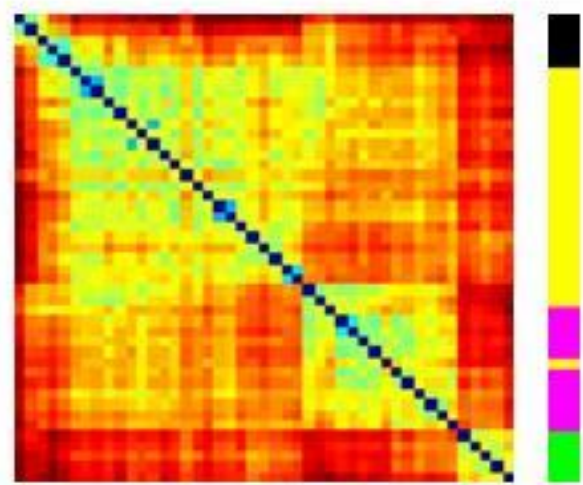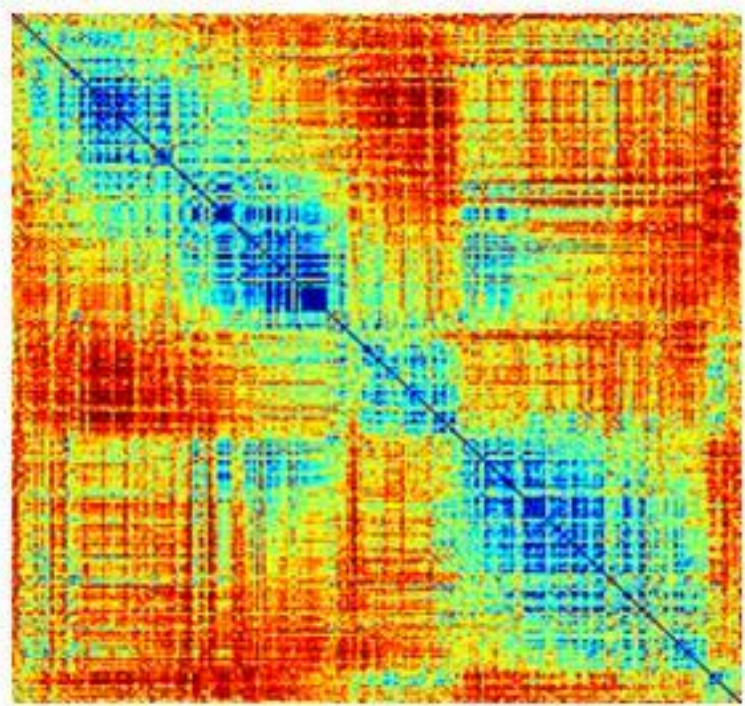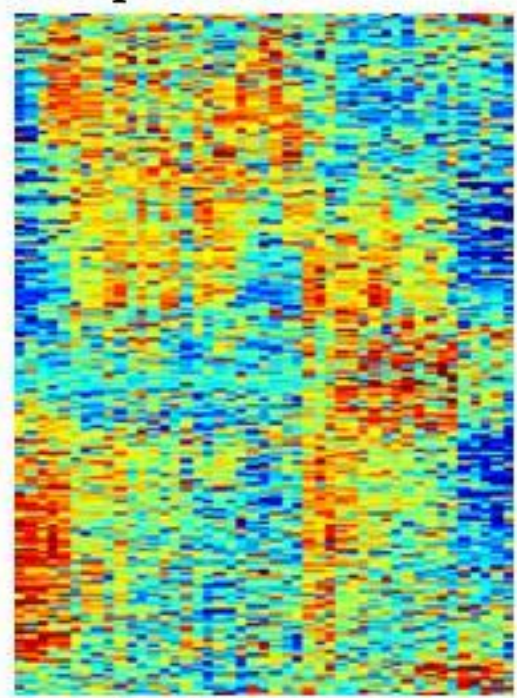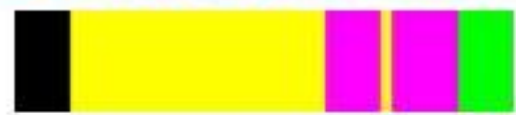

**a.**

**b.**

**c.**

**d.**

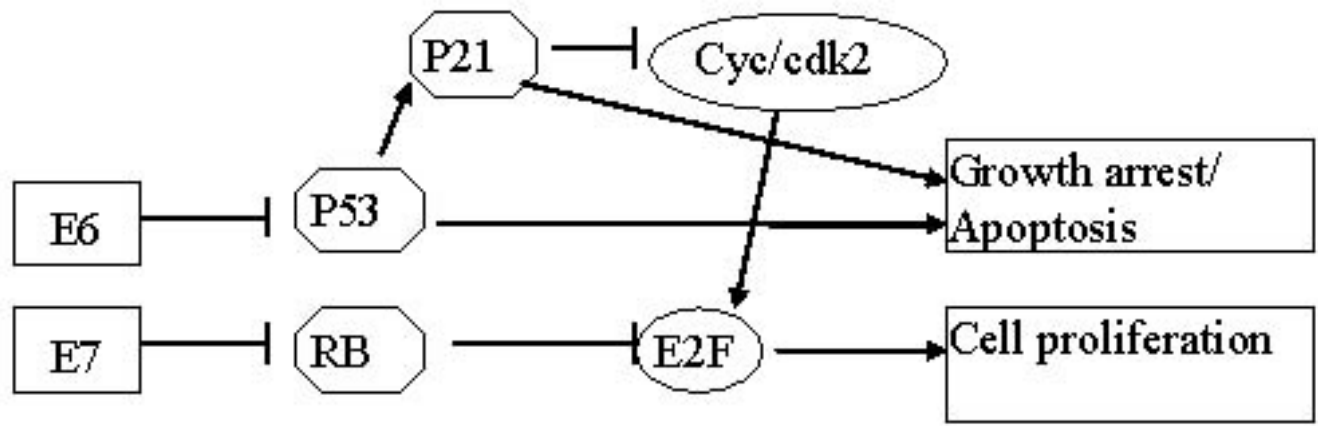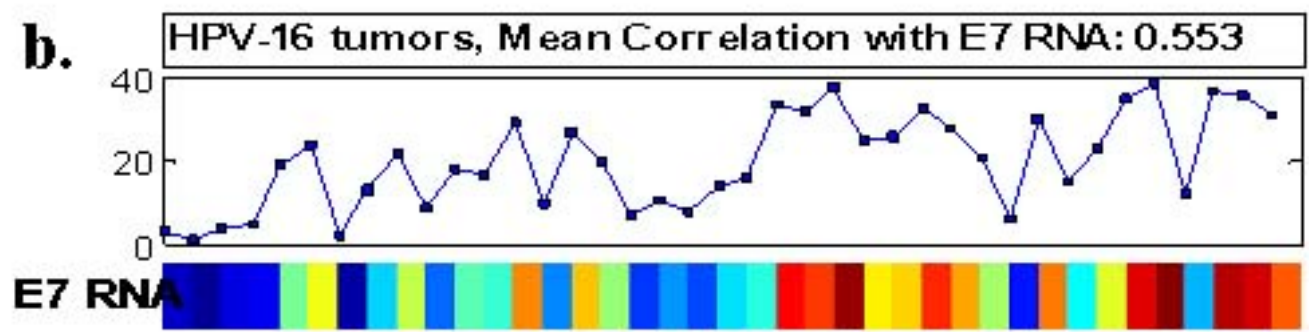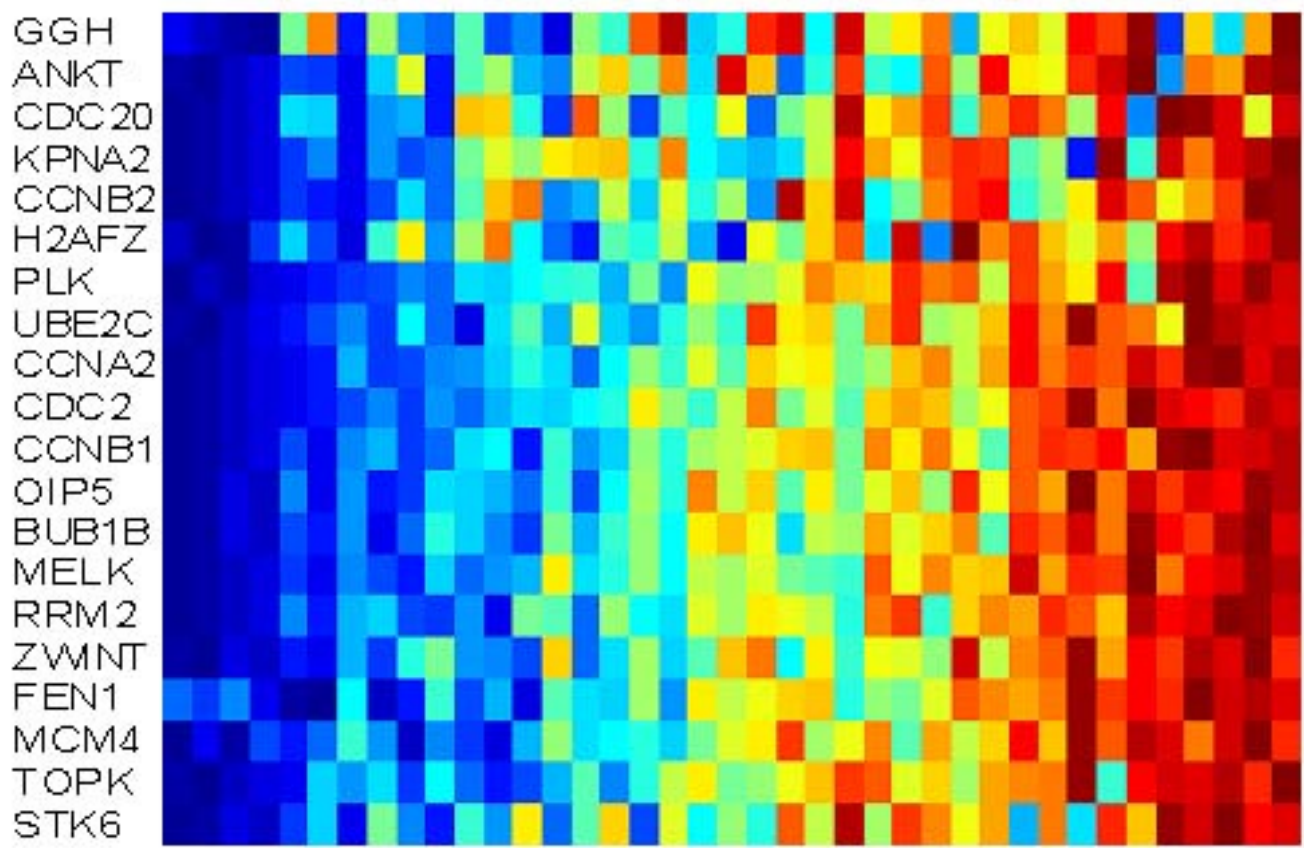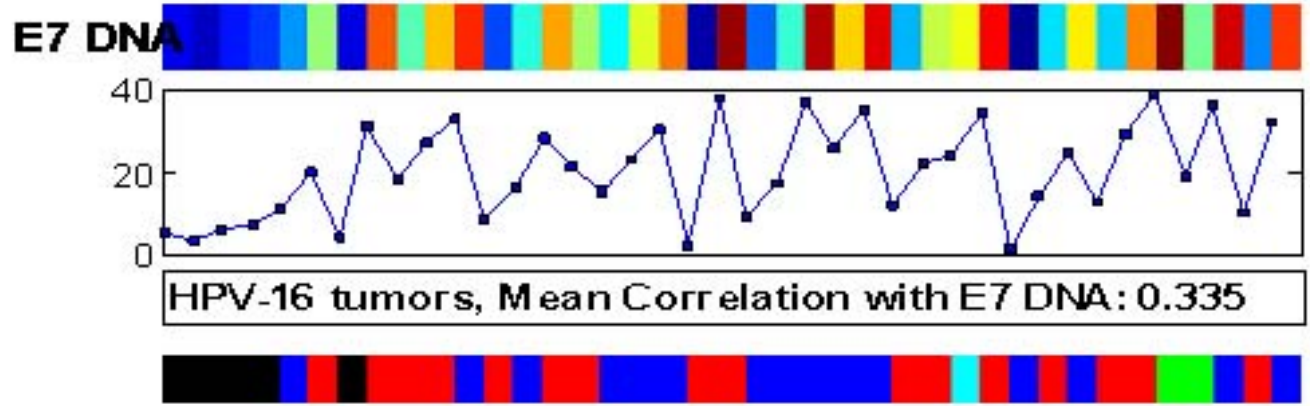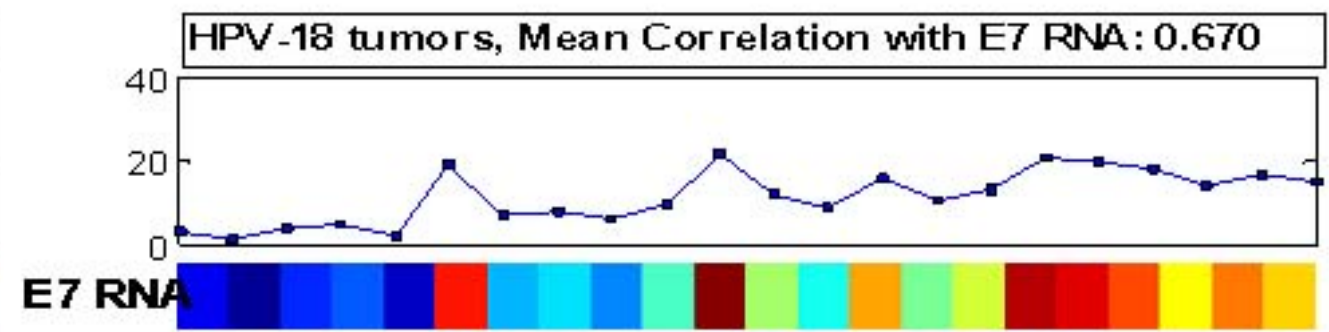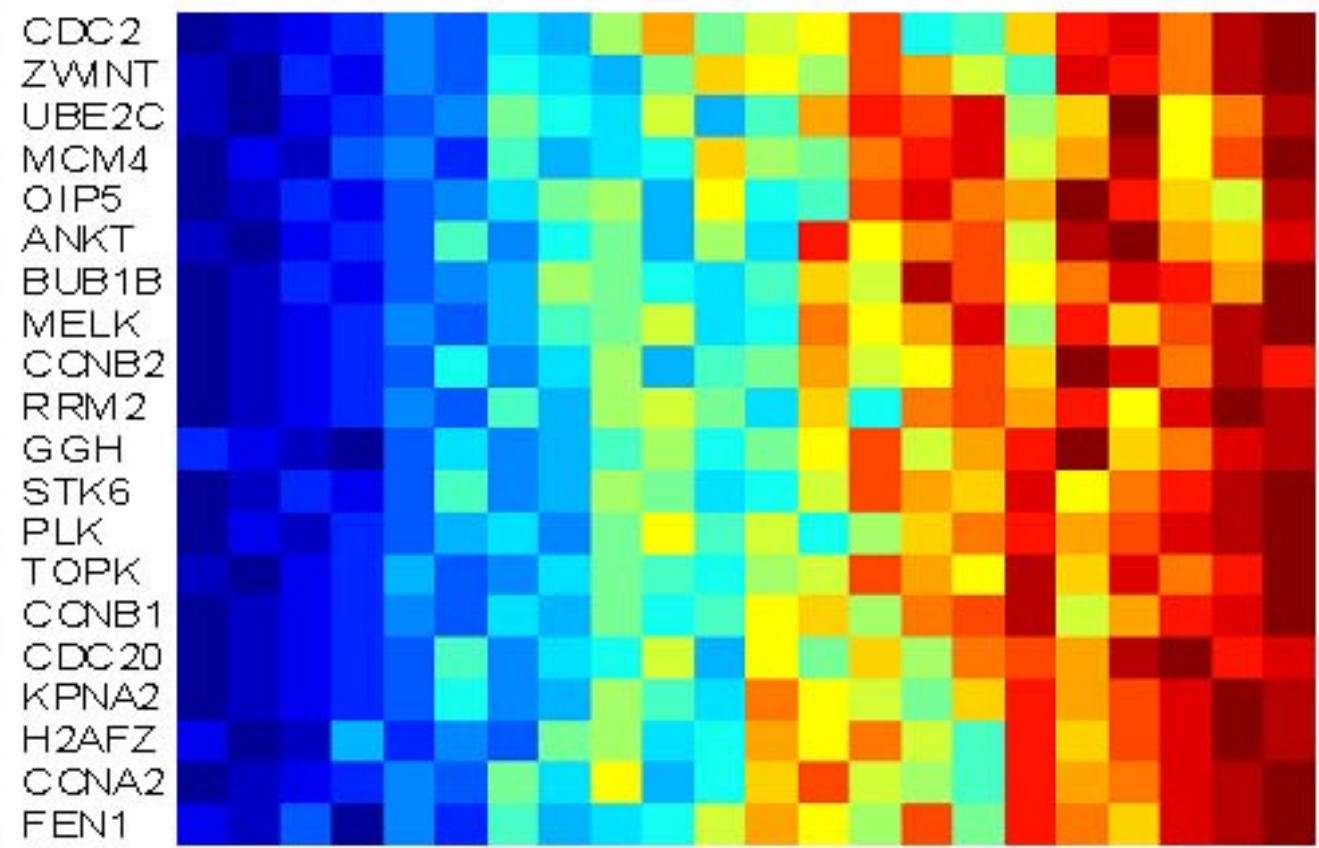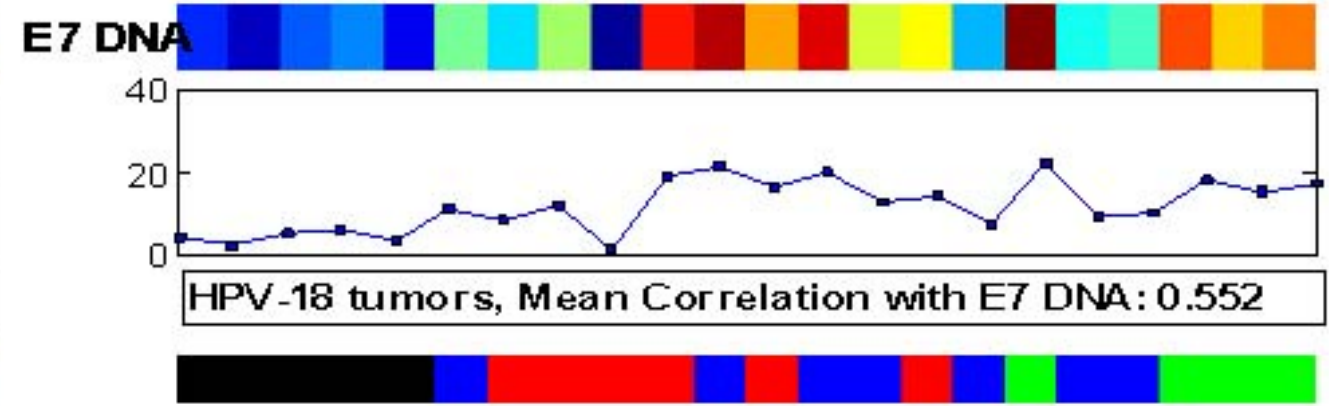